%% file: discoptdev.tex
\documentclass[twocolumn,showpacs,pra,aps,superscriptaddress]{revtex4}
\usepackage[dvips]{graphicx}
\usepackage{amsthm,amsmath,amssymb}
\usepackage{bbold}
\input{myqcircuit}

\newcommand{\braket}[2]{\langle#1|#2\rangle}
\newcommand{\ketbra}[2]{\ket{#1}\bra{#1}}

\newcommand{\Tr}{\operatorname{Tr}}
\newcommand{\re}{\operatorname{Re}}

\newcommand{\der}[2]{\frac{\partial #1}{\partial #2}}
\newcommand{\floor}[1]{\lfloor#1\rfloor}
\newcommand{\ceil}[1]{\lceil#1\rceil}
\newcommand{\hilb}[1]{\mathcal{#1}}

\newtheorem{lmm}{Lemma}
\newtheorem{thm}{Proposition}

\begin{document}
\title{Tradeoff between energy and error in the discrimination of quantum-optical devices}

\author{Alessandro \surname{Bisio}}
\author{Michele \surname{Dall'Arno}}
\author{Giacomo Mauro \surname{D'Ariano}}

\affiliation{Quit group, Dipartimento di Fisica ``A. Volta'', via Bassi 6,
  27100 Pavia, Italy}
\affiliation{Istituto Nazionale di Fisica Nucleare,
  Gruppo IV, via Bassi 6, 27100 Pavia, Italy}

\date{\today}

\begin{abstract}
  We address the problem of energy-error tradeoff in the discrimination
  between two linear passive quantum optical devices with a single use. We
  provide an analytical derivation of the optimal strategy for beamsplitters
  and an iterative algorithm converging to the optimum in the general case. We
  then compare the optimal strategy with a simpler strategy using coherent
  input states and homodyne detection. It turns out that the former requires
  much less energy in order to achieve the same performances.
\end{abstract}

\maketitle

\section{Introduction}

There are many contexts in which one wishes to use as little energy as
possible in order to perform a desired task. Consider the case in which one
wants to gather information about an unknown device. The most general approach
to the problem is to probe the target and to measure its response. This
implies an undesired perturbation of the system. Often such perturbation
depends on the energy of the probe, and one can ask what is the minimum amount
of energy the probe must have in order to perform the desired task. Exploiting
the distinctive features of quantum theory, such as entanglement, one could in
principle achieve better performances with respect to a classical strategy
with the same energy. For example, in the context of quantum illumination
\cite{Lloydscience, TEGGLMPS08, Lloydshapnewjphys}, one discovers that the use
of entangled light allows to enhance the detection and the imaging of an
object.

The task we are considering in this paper is the discrimination between two
quantum optical devices (such as beamsplitters or phase shifters). An optical
device is represented by a unitary transformation $U$ acting on a system
described by a Fock space $\hilb{H}$. The general problem in which one wants
to discriminate between two unitaries has been widely investigated in the
literature \cite{Aci01,DLP02,DFY07, sedlakdiscrmin, collinsdiscrimin} and can
be summarized as follows. Suppose we are provided with a single use
\cite{note} of an unknown device randomly chosen in a set of two $U_1$ and
$U_2$, with equal prior probability. Our task is to distinguish between $U_1$
and $U_2$ with a probability of error not greater than a given threshold
$q$. One sends an input state $\ket{\psi}$ that undergoes the unitary
evolution described either by $U_1$ or $U_2$. Finally, one applies the optimal
POVM \cite{Hel76} in order to distinguish the two possible output states
$U_1\ket{\psi}$ or $U_2\ket{\psi}$ with minimum error probability. This error
probability depends on the input $\ket{\psi}$, so one can optimize on
$\ket{\psi}$. In the case at hand, $U_1$ and $U_2$ describe optical devices
and the input state $\ket{\psi}$ is a quantum state of light, with a
well-defined mean value of the energy. Since in general many states allow to
discriminate with an error probability $P_e \le q$, one can wonder which state
$\ket{\psi^*}$ accomplishes this task with minimum mean value of the energy.

Here we analyze the energy-error tradeoff in the case in which the optical
devices to be discriminated are linear, passive and lossless. This
discrimination is useful for example when reading classical digital
information encoded in the reflectivity of a media, such as conventional CDs
or DVDs. This particular application has first been suggested in \cite{Pir11}
for the discrimination of quantum channels affected by loss and noise. In
particular, there it is shown that, for fixed mean number of photons
irradiated, non-classical light can outperform any classical source in terms of
the amount of information retrieved. Recently, the discrimination of lossy
beamsplitters has been considered in \cite{Nai11}, in the scenario where one
has access to only one of the input and output modes and to a restricted class
of input states. A similar scenario, namely the discrimination of a lossy
quantum channel from an ideal one, has been studied also in \cite{IPP10}.

The paper is organized as follows. In Section \ref{sec:discr-line-pass} we
review some basic notions of linear optics and introduce the problem. The
analytical derivation of the energy-error tradeoff when the devices are
beamsplitters is given in Section \ref{sec:case-bemsplitters}, while a
numerical algorithm for the solution in the general case is presented in
Section \ref{sec:an-iter-algor}.  In Section \ref{sec:discr-pass-devic} we
analyze the energy-error tradeoff in the restricted scenario in which only
coherent states and homodyne detections are available. We will then quantify
the advantages that one has by adopting the optimal quantum strategy. Section
\ref{sec:conclusion} concludes the paper with a discussion of the results.

\section{Discrimination of linear passive optical devices}
\label{sec:discr-line-pass}

A $M$-modes quantum optical device \cite{VW06, leonhardt03} is described by a
unitary operator $U$ relating $M$ input optical modes with annihilation
operators $a_i$ on $\hilb{H}_i$, to $M$ output optical modes with annihilation
operators $a_i'$ on $\hilb{H}_{i'}$, where $\hilb{H}_i$ denotes the Fock space
of the optical mode $i$. We denote as $\hilb{H}=\bigotimes_i \hilb{H}_i$.

An optical device is called \emph{linear} if the operators of the output modes
are related to the operator of the input modes by a linear transformation,
namely
\begin{equation*}
  a' := U a_i U^\dagger = \sum_{j=1}^N A_{ij} a_j + \sum_{j=1}^N B_{ij}
  a^\dagger_j, \quad i =1, \dots, N.
\end{equation*}
The above equation can be rewritten in the more compact form
\begin{align}
  \left(\begin{array}{c}{\bf a'}\\{\bf a'}^\dagger\end{array}\right) = S
    \left(\begin{array}{c}{\bf a}\\{\bf a}^\dagger\end{array}\right),
\end{align}
where $S$ is the $2N\times2N$ scattering matrix defined as
\begin{align}\label{eq:smatrix}
  S := \left( \begin{array}{cc} A & B \\ \bar{B}&\bar{A} \end{array} \right)
\end{align}
($\bar{X}$ is the complex conjugate of $X$), ${\bf a} = (a_1,\dots a_N)$ is
the vector of annihilation operators of the input mode, and analogously
${\bf a'}$ for the output modes. If $B=0$ in Eq. \eqref{eq:smatrix} the device
is called \emph{passive} and conserves the total number of photons, that is
\begin{align}\label{eq:passivedevconsen}
  \bra{\psi} N \ket{\psi} = \bra{\psi} U^\dagger N U\ket{\psi}
\end{align}
with $N := \sum_i a_i^\dagger a_i$ the \emph{number operator} on $\hilb{H}$.

Suppose now that we want to discriminate between two linear optical passive
devices $U_1$ and $U_2$. If a single use of the unknown device is available,
the most general strategy consists of i) preparing a bipartite input state
$\rho \in \mathcal{B}(\hilb{H} \otimes \hilb{K})$ ($\hilb{K}$ is an ancillary
Fock space with mode operators $b_i$), ii) applying locally the unknown device
and iii) performing a bipartite POVM ${\Pi} = \{ \Pi_1, \Pi_2 \}$ on the
output state $(\mathcal{U}_x \otimes \mathcal{I}_{\mathcal{K}} ) \rho = (U_x
\otimes I_\mathcal{K})\rho (U^\dagger_x \otimes I_\mathcal{K}) $ ($x $ can be
either $1$ or $2$)
\begin{align}
  \begin{aligned}
    \Qcircuit @C=0.7em @R=1em { \multiprepareC{1}{\rho} & \ustick{\mathcal{H}}
      \qw & \gate{U_x} & \qw & \ghost{\Pi}\\ \pureghost{\rho} &
      \ustick{\mathcal{K}} \qw & \qw & \qw & \multimeasureD{-1}{\Pi} }
  \end{aligned}\quad .
\end{align}
When the device is randomly chosen from the set $ \{U_1, U_2 \}$ with equal
probabilities $p_1=p_2=1/2$, the minimum probability of error in the
discrimination can be proved to be \cite{NC00}
\begin{align}\label{eq:error1}
  P_e (\rho, U_1, U_2) = \frac12 \left( 1- ||(\mathcal{U}_1 \otimes
  \mathcal{I}_{\mathcal{K}} ) \rho -(\mathcal{U}_2 \otimes
  \mathcal{I}_{\mathcal{K}} ) \rho ||_1 \right).
\end{align}
where $|| X ||_1 = \Tr[\sqrt{X^\dagger X}]$. If we define as $N_{\mathcal{K}}
= \sum_i b^\dagger _i b_i$ the number operator on the ancillary modes $b_i$,
the energy of the state $\rho$ is proportional to
\begin{align}\label{eq:energy}
  E(\rho) := \frac12 + \Tr[\rho(N \otimes I_{\mathcal{K}} + I \otimes
    N_{\mathcal{K}} )],
\end{align}
while clearly the energy that flows through the unknown device is $E_d(\rho)
:= \frac12 + \Tr[\rho (N \otimes I_{\mathcal{K}})]$.

Since we have $P_e((\mathcal{U}^\dagger_1 \otimes
\mathcal{I}_{\mathcal{K}})\rho, I, U_2U_1^\dagger ) = P_e(\rho, U_1, U_2)$ and
Eq. \eqref{eq:passivedevconsen} implies $E((\mathcal{U}^\dagger_1 \otimes
\mathcal{I}_{\mathcal{K}})\rho) = E(\rho)$, we can restrict our analysis to
the case in which $U_1 = I$ and $U_2 = U$.  We consider now the problem to
find the minimum energy input state $\rho^*$ that allows us to discriminate
between $ I$ and $U$ with probability of error not greater than a given
threshold $q$, that is
\begin{align}\label{eq:figmer1}
  \rho^* = \arg \min_{\rho \textrm{ s.t. } P_e(\rho) \leq q} E(\rho).
\end{align}
Sometimes one is more interested in minimizing the energy $E_d(\rho)$ flowing
through the device, rather than the total energy $E(\rho)$ of the input
state. In the following we show that no ancillary modes are required for
optimal discrimination, so these minimization problems lead to the same
optimal input state.

The following lemmas allow us to simplify Eq. \eqref{eq:figmer1}.
\begin{lmm}\label{lmm:wall}
  Without loss of generality the minimization in Eq. \eqref{eq:figmer1} can
  be rewritten as
  \begin{align}
    \rho^* = \arg \min_{\rho \textrm{ s.t. } P_e(\rho,U) = q} E(\rho).
  \end{align}
\end{lmm}

\begin{proof}
  Suppose that for the optimal state $\rho^*$ one has $P_e(\rho)
  < q$.  Since $P_e(\rho)$ is a continuous function there exists a $0
  < \alpha \leq 1$ such that $P_e((1 - \alpha )\rho + \alpha
  \ketbra{0}{0} ) = q$, where $\ketbra{0}{0}$ denotes the vacuum
  state.  By observing that $E((1 - \alpha )\rho + \alpha
  \ketbra{0}{0}) < E(\rho)$ we have the thesis.
\end{proof}

\begin{lmm}\label{lmm:pure}
  The optimal state achieving the minimum in Eq. \eqref{eq:figmer1} can be
  chosen pure.
\end{lmm}

\begin{proof}
  First, let us prove that the minimization of $E(\rho)$ for a given value of
  $P_e$ is equivalent to the minimization of the convex combination
  \begin{align}\label{eq:figmer2}
    F(\rho) = p P_e(\rho) + (1-p) E(\rho)
  \end{align}
  for a fixed value of $p$.  Suppose that we find $\rho^*$that minimizes
  $F(\rho)$. It is then clear that for $q := P_e(\rho^*)$, $E(\rho)$ gives the
  minimum possible value for the energy because any lower energy would
  decrease $F$. From Eq. \eqref{eq:error1} and the convexity of the trace
  distance, it follows that $P_e(\rho)$ is a concave function of $\rho$.
  Since $P_e(\rho)$ is concave and $E(\rho)$ is linear, $F(\rho)$ is a concave
  function of $\rho$ and its minimum is attained on the boundary of its
  dominion, i. e. for pure states.
\end{proof}
Lemma \ref{lmm:pure} allows us to rewrite Eq. \eqref{eq:error1} as
\begin{align}\label{eq:error2}
  P_e = \frac12 \left( 1- \sqrt{1-|\bra{\psi}(U \otimes
    I_{\mathcal{K}})\ket{\psi} |^2}\right).
\end{align}

We now prove that no ancillary modes are required.
\begin{lmm}
  Without loss of generality, the minimum in \eqref{eq:figmer1} is achieved
  without using ancillary modes.
\end{lmm}

\begin{proof}
  Due to Lemma \ref{lmm:pure} the input state can be written as $ \ket{\psi} =
  \sum_i c_i \ket{i}\ket{\chi_i} $ where $\ket{i}$ is an orthonormal basis in
  $\mathcal{H}$ and $\ket{\chi_i}$ are normalized states in $\mathcal{K}$.  If
  we define $\ket{{\psi}'} := \sum_i c_i \ket{i}\ket{0}$ ($\ket{0}$ is the
  vacuum state), it is easy to verify that $P_e(\psi) = P_e({\psi}')$ while
  $E(\psi) \geq E({\psi}') $.
\end{proof}

After these consideration and disregarding some irrelevant constant factors,
Eq. \eqref{eq:figmer1} can be rewritten as
\begin{align}\label{eq:figmer3}
  \ket{\psi^*} = \arg\min_{\ket{\psi} \textrm{s.t.} |\bra{\psi} U \ket{\psi}|
    = K} \bra{\psi}N\ket{\psi},
\end{align}
where from Eq. \eqref{eq:error2} it follows that $K=\sqrt{4q(1-q)}$.

\section{The Optimal tradeoff}

\subsection{The case of beamsplitters}
\label{sec:case-bemsplitters}

In this section we derive an analytical expression for the optimal
energy-error tradeoff for the case in which the devices to be
discriminated are two beamsplitters.
A beamsplitter is a two-mode linear passive quantum optical device
whose scattering matrix $S$ has the form
\begin{align}\label{eq:smatrix2}
  S := \left( \begin{array}{cc} A & 0 \\ 0 & \bar{A} \end{array} \right)\qquad
  A \in SU(2).
\end{align}
In the following we will use the basis $\{ \ket{nm} \}$ with respect to which
$U$ is diagonal i.e.
\begin{align}\label{eq:su2matrix}
  A = \begin{pmatrix}
    e^{i \delta} & 0\\ 
    0 & e^{- i \delta}
  \end{pmatrix},
  \qquad 0 \leq \delta \leq \pi
\end{align}
With this choice, for any $\ket{\psi} = \sum_{n,m=0}^\infty \alpha_{nm}
\ket{n,m}$, we have $U \ket{nm} = e^{i\delta (n-m)}\ket{nm}$ and it is easy to
observe that $ \bra{\psi} U \ket{\psi} = \sum_{n,m=0}^\infty |\alpha_{nm}|^2
e^{i\delta(n-m)} $ and $\bra{\psi}N\ket{\psi} = \sum_{n,m=0}^\infty
|\alpha_{nm}|^2 (n+m)$. We notice that both these expressions only depends on
the squared modulus of the coefficients $\alpha_{nm}$ and so we can assume
$\alpha_{nm}$ to be real and positive. The assumption that the devices are
beamsplitters allows us to simplify the structure of the optimal input state
for the tradeoff. First, we show that it is not restrictive to consider
superpositions of the so called NOON \cite{San89} states.
\begin{lmm}\label{thm:symmetry}
  If $U$ is a beamsplitter, the optimal state
  $\ket{\psi^*}$ in Eq. \eqref{eq:figmer3} can be taken of the form
  \begin{align}\label{eq:state}
    \ket{\psi^*} = \sum_{n=0}^{\infty} \alpha_n \ket{\phi_n}, \quad
    \ket{\phi_n} = \sqrt{\frac12}(\ket{n,0}+\ket{0,n})
  \end{align}
\end{lmm}

\begin{proof}
  We show that for any state $\ket{\psi} = \sum_{n,m=0}^\infty \alpha_{nm}
  \ket{n,m}$ there exists a state $\ket{\psi'} = \sum_{l=0}^\infty \alpha'_{l}
  \ket{\phi_l}$ such that $\bra{\psi'}N\ket{\psi'} \le \bra{\psi}N\ket{\psi}$
  and $|\bra{\psi'}U\ket{\psi'}| \le |\bra{\psi}U\ket{\psi}|$.  Upon defining
  $|\alpha'_l|^2 = \sum_{|n-m|=l} |\alpha_{nm}|^2$, one can verify
  \begin{align}
    \bra{\psi'}N\ket{\psi'} & = \sum_{n,m=0}^\infty \alpha_{nm}^2 |n-m| \le
    \bra{\psi}N\ket{\psi},\\ | \bra{\psi'}U\ket{\psi'} | & = \left |
    \sum_{n,m=0}^\infty \alpha_{nm}^2 \cos(\delta|n-m|) \right| \le \\& \le
    |\bra{\psi}U\ket{\psi} |, \nonumber
  \end{align}
  that proves the statement.
\end{proof}

From Eq. \eqref{eq:state} it follows that the expectation value of $U$ over
$\ket{\psi}$ is real. Following an argument similar to the one we used to
prove Lemma \ref{lmm:wall} , the constraint $|\bra{\psi} U \ket{\psi}| = K$
can be changed into $\bra{\psi} U \ket{\psi} = K$ and Eq. \eqref{eq:figmer3}
becomes
\begin{align}\label{eq:figmer5}
  \ket{\psi^*} = \arg\min_{\ket{\psi}s.t. \bra{\psi} U \ket{\psi} = K}
  \bra{\psi}N\ket{\psi}.
\end{align}
One can observe that for any state $\ket{\psi}$ of the form of
Eq. \eqref{eq:state}, the constraint $ \sum_{n=0}^\infty \alpha_n^2
\cos(\delta n) = K$ implies that there must exist at least one non-null
coefficient $\alpha_n$ for $n$ s.t. $\cos(\delta n)\le K$ and at least one for
$n$ s.t. $\cos(\delta n) \ge K$. In the following lemma we will prove that not
more than two non-null coefficient $\alpha_n$ are needed.

\begin{lmm}\label{thm:trapezium}
  The optimal state $\ket{\psi^*}$ that minimize Eq. \eqref{eq:figmer5} can be
  taken of the form
  \begin{align}\label{eq:discstate}
    \ket{\psi^*} = \alpha_{n_1}\ket{\phi_{n_1}} + \alpha_{n_2}
    \ket{\phi_{n_2}},
  \end{align}
\end{lmm}

\begin{proof}
  Let us consider the optimal state $\ket{\psi} = \sum_n \alpha_{n}
  \ket{\phi_n}$ with $\bra{\psi}U\ket{\psi}=K$ and
  $\bra{\psi}N\ket{\psi}=N_{\min}$. Suppose now that the set ${ \vec{\alpha}}
  := \{\alpha_{n} \}$ has $N \geq 3$ elements. Then there must exist $n_1$ and
  $n_2$ such that $\alpha_{n_1},\alpha_{n_2}\neq 0$ and $\cos(\delta n_1) \leq
  K \leq \cos(\delta n_2)$. It is then possible to define $\ket{\chi}:=
  \beta_{n_1}\ket{\phi_{n_1}} + \beta_{n_2}\ket{\phi_{n_2}}$ such that
  $\bra{\chi}U\ket{\chi}=K$. Furthermore, we can define $\ket{\xi} :=
  (1-\epsilon)^{-1/2}\sum_n \gamma_n \ket{\phi_n}$, where
  \begin{align*}
   \gamma_n = \left\{ \begin{array}{ll} \alpha_n & \mbox{ if } n \ne
        n_1,n_2\\ \sqrt{\alpha_n^2 - \epsilon \beta_n^2} & \mbox{ if } n =
        n_1,n_2  \end{array} \right. ,
  \end{align*}
  and $\epsilon \le
  \min(\alpha_{n_1}/\beta_{n_1},\alpha_{n_2}/\beta_{n_2})$.
  We notice that $\bra{\xi}U\ket{\xi}=K$, and
  \begin{align}
    N_{\min} = \epsilon \bra{\chi}N\ket{\chi} +
    (1-\epsilon) \bra{\xi}N\ket{\xi}.
  \end{align}
  If $\bra{\chi} N \ket{\chi} = N_{\min}$ the statement follows with
  $\ket{\psi} = \ket{\chi}$. If $\bra{\chi}N\ket{\chi} \neq N_{\min}$, either
  $\bra{\chi}N\ket{\chi} < N_{\min}$ or $\bra{\xi}N\ket{\xi} < N_{\min}$, that
  contradicts the hypothesis that $\ket{\psi}$ is the optimal state.
\end{proof}

Since we both require $\bra{\psi}U\ket{\psi} = K$ and $\alpha_{n_1}^2 +
\alpha_{n_2}^2 = 1$ the expression of the coefficients $\alpha_{n_1},
\alpha_{n_2}$ and of the mean value $\bra{\psi}N\ket{\psi}$ are fixed by the
choices of $n_1$, $n_2$ and $K$, i.e.
\begin{align}\label{eq:energy2}
  \alpha_{n_1} = \sqrt{ \frac {\cos(\delta n_2)-K} {\cos(\delta
      n_2)-\cos(\delta n_1)}} ,\quad
  \alpha_{n_2} = \sqrt{1-\alpha_{n_1}^2 } \\
  \bra{\psi}N\ket{\psi} = \frac{n_2 \cos(\delta n_1) - n_1\cos(\delta n_2) +
    K(n_1-n_2)}{\cos(\delta n_1)-\cos(\delta n_2)}.
\end{align}

\begin{figure}[htb]
  \includegraphics{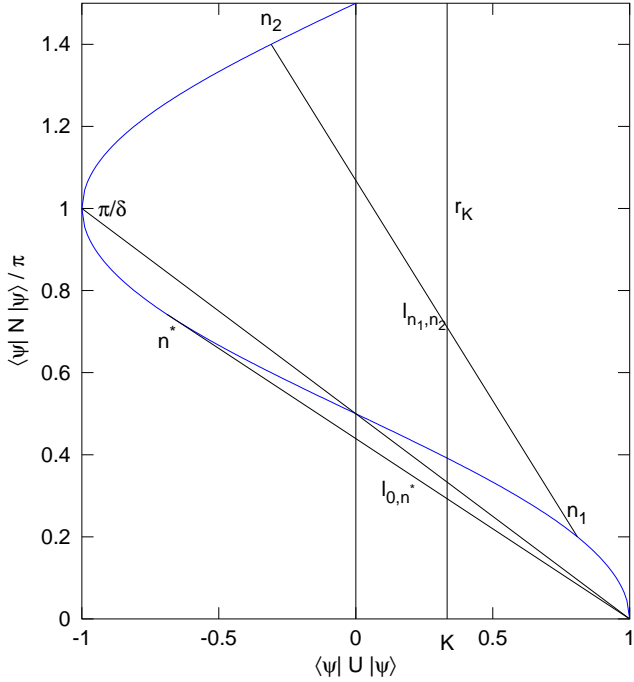}
  \caption{(Color online) Geometrical representation of the optimization
    problem. For fixed values $n_1$ and $n_2$, each state of the form
    \eqref{eq:discstate} is represented by a point of the line segment
    $l_{n_1,n_2}$ which extrema are $(cos(\delta n_1),n_1)$ and $(cos(\delta
    n_2),n_2)$. The x and y coordinates represent the expectation values
    $\bra{\psi}U\ket{\psi}$ and $\bra{\psi}N\ket{\psi}$, respectively.  From
    the picture it is clear that the line segment $l_{0,n^*}$ lies below any
    other line segment $l_{n_1,n_2}$, and so it identifies the family of
    optimal states, as we prove in Prop. \ref{thm:beamopt}.}
  \label{fig:arccos}
\end{figure}

It is now convenient to rephrase the problem at hand in a geometrical way (see
Fig. \ref{fig:arccos}). Let us introduce the map $\mathsf{f}: \mathbb{N} \to
\mathbb{R}^2$, defined as $\mathsf{f}(n) = (\cos(\delta n),n)$.  We can
associate to any couple $n_1$ and $n_2$ the line $L_{n_1, n_2} := \{ t \,
\mathsf{f}(n_1) + (1-t) \mathsf{f}(n_2) | t \in \mathbb{R} \}$ and the segment
$l_{n_1, n_2} := \{ t \, \mathsf{f}(n_1) + (1-t) \mathsf{f}(n_2) | 0 \leq t
\leq 1 \}$. It is easy to prove that the expectation value
$\bra{\psi}N\ket{\psi}$ in Eq. \eqref{eq:energy2} can be rewritten as
\begin{align}
  \bra{\psi}N\ket{\psi} = (\ell_{n_1, n_2} \cap r_K)_y
\end{align}
where we defined the line $r_K := \{ (K,y)| y \in \mathbb{R} \}$ and the
mapping $(a,b)_y = b$. For any $K$ it is then possible to define a partial
ordering $<_K$ among the segments $l_{n_1,n_2}$ for which $\cos (\delta n_1)
\le K \le \cos (\delta n_2)$, as follows:
\begin{align}
  l_{n_1,n_2} >_K l_{n'_1,n'_2} \textrm{ if } (l_{n_1,n_2} \cap r_K)_y >
  (l_{n'_1,n'_2} \cap r_K)_y
\end{align}

We are now ready to proof the following Proposition. Here $\floor{x}
(\ceil{x})$ denotes the minimum (maximum) integer number greater (smaller)
than $x$.
\begin{thm}\label{thm:beamopt}
  For any $K$, $0 \leq K \leq 1$, the optimal state $\ket{\psi^*}$ that
  minimizes Eq. \eqref{eq:figmer5} is
  \begin{align}\label{optimalstate}
    \ket{\psi^*} = \alpha_{n^*} \ket{\phi_{n^*}} + \alpha_{0} \ket{00},
  \end{align}
  where $n^* = \arg \min_{\floor{\tilde{n}},\ceil{\tilde{n}}} \bra{\psi^*} N
  \ket{\psi^*}$, with $\tilde{n}$ the minimum positive solution of $\delta n =
  \tan (\delta n/2)$.
\end{thm}

\begin{proof}
  First let us introduce the set $\Omega := \{n \in \mathbb{N} | \pi/2\delta
  \le n' \le \pi/\delta \}$.  Since $\delta < \pi$ we observe that $\Omega$
  must be nonempty. Consider now the lines $L_{n',0}$ and $L_{n',m}$. It is
  easy to verify that for all $m$ we have $ (L_{n',m} \cap r_1)_y > (L_{n',0}
  \cap r_1)_y$ which implies $ (L_{n',m} \cap r_K)_y > (L_{n',0} \cap r_K)_y$
  for all $\cos(\delta n') \leq K \leq 1 $ and finally $l_{n',m} >_K l_{n',0}$
  for all $\cos(\delta n') \leq K \leq \cos(\delta m)$. Similarly one can
  verify the bound $ l_{n',0} <_K l_{\pi/\delta , 0} <_K l_{n_1,n_2} $ that
  holds for $\cos(\delta n_1) \leq K \leq \cos(\delta n_2)$, $n' \in \Omega$.
  We can then restrict the optimization over the finite set of states $S :=
  \{\ket{\psi}| \ket{\psi} = \alpha_{n'}\ket{\phi_{n'}} + \alpha_{0}\ket{00},
  n' \in \Omega \}$. As a consequence, the mean value $\bra{\psi} N
  \ket{\psi}$ becomes
  \begin{align}\label{eq:energy3}
 \bra{\psi} N \ket{\psi}   = \frac{(1-K) n}{1-\cos(\delta n)}.
  \end{align}
  The right hand side of Eq. \eqref{eq:energy3} can be proven to be a convex
  function \cite{BV04} for $\pi/2 \leq \delta n \leq \pi$ and achieves its
  minimum for $\tilde{n}$ minimum positive solution of $\delta n = \tan
  (\delta n/2)$. Since $\tilde{n}$ is in general not integer the optimal value
  $n^*$ is given by $\arg \min_{\floor{\tilde{n}},\ceil{\tilde{n}}}
  \bra{\psi^*} N \ket{\psi^*}$.
\end{proof}

\begin{figure}[htb]
  \includegraphics{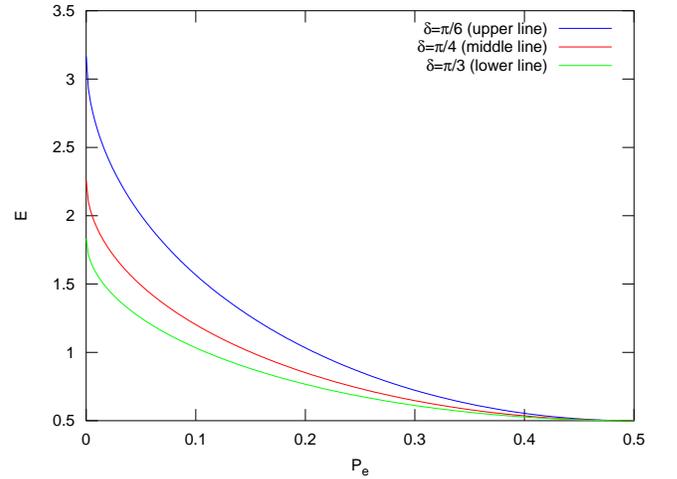}
  \caption{(Color online) Optimal tradeoff between the energy $E$ and the
    probability of error $P_e$ in the discrimination of I and $U=\exp(i(\delta
    a_1^\dagger a_1-\delta a_2^\dagger a_2))$, for various values of
    $\delta$.}
  \label{fig:su2}
\end{figure}

Figure \ref{fig:su2} shows the optimal energy-error tradeoff obtained with the
discrimination strategy of Prop. \ref{thm:beamopt}.

\subsection{An iterative algorithm for the general case}
\label{sec:an-iter-algor}

Here we provide an iterative steepest-descent algorithm \cite{Arf85} to find a
state $\ket{\psi}$ that solves the optimization problem in
Eq. \eqref{eq:figmer3}.  With the same argument we use in the proof of Lemma
\ref{lmm:pure} we can rephrase the optimization problem in
Eq. \eqref{eq:figmer3} as the minimization of the convex combination
\begin{align}\label{eq:figmer4}
  C(\psi) := p \bra{\psi}N\ket{\psi} + (1-p) |\bra{\psi}U \ket{\psi}|^2 .
\end{align}

We are now ready to introduce the iterative procedure.
\begin{thm}\label{thm:algorithm}
  The following algorithm converges to a state $\ket{\psi}$ that is optimal
  according to Eq. \eqref{eq:figmer4}. Take an arbitrary state $\ket{\psi^0}$.
  Given $\ket{\psi^n}$, evaluate $\ket{\psi^{n+1}}$ by the following steps:
  \begin{enumerate}
    \item Evaluate the derivative of the figure of merit $C(\psi^n)$
      \begin{align}
        \der{C(\psi^n)}{\bra{\psi^n}} = [ & p N + (1-p) (
          \bra{\psi^n}U\ket{\psi^n} U^\dagger +\\ & +
          \bra{\psi^n}U^\dagger\ket{\psi^n} U ) ] \ket{\psi^n}.
      \end{align}
    \item Pick up a small enough positive $\alpha$ and evaluate
      \begin{align}
        \ket{\hat{\psi}^{n+1}} = (1-\alpha)\ket{\psi^n} - \alpha
        \der{C(\psi^n)}{\bra{\psi^n}}.
      \end{align}
    \item Normalize $\ket{\hat{\psi}^{n+1}}$ according to
      \begin{align}
        \ket{\psi^{n+1}} = |\hat{\psi}^{n+1}|^{-1} \ket{\hat{\psi}^{n+1}}.
      \end{align}
  \end{enumerate}
\end{thm}

\begin{proof}
  The algorithm in Prop. \ref{thm:algorithm} is a steepest-descent algorithm:
  we move the state in the direction of the gradient of the figure of merit
  $C(\psi)$, so by construction one has $C(\psi^{n+1}) \le C(\psi^n)$.
\end{proof}

\begin{figure}[htb]
  \includegraphics{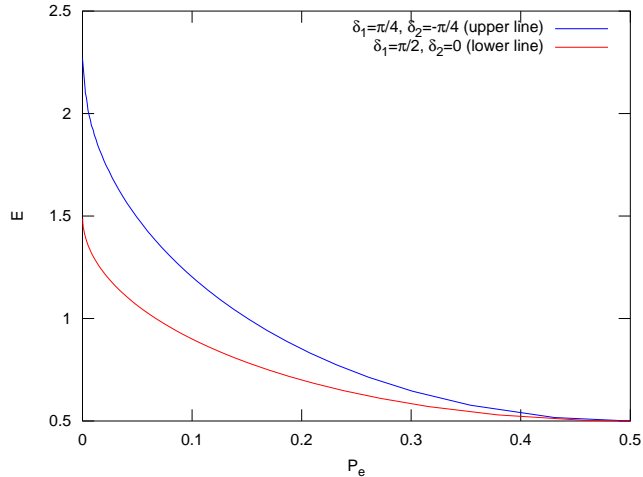}
  \caption{(Color online) Optimal tradeoff between the energy $E$ and the
    probability of error $P_e$ in the discrimination of two linear passive
    optical devices $I$ and $U$. We consider the case in which $U$ is a two
    mode device, namely $U=\exp(i(\delta_1 a_1^\dagger a_1 + \delta_2
    a_2^\dagger a_2))$.}
  \label{fig:u2}
\end{figure}

The parameter $\alpha$ controls the length of each iterative step, so for
$\alpha$ too large an overshooting can occur. This can be kept under control
by evaluating the figure of merit $C(\psi)$ at the end of each step: if
$C(\psi)$ increases instead of decreasing, we are warned that we have taken
$\alpha$ too large.  Figure \ref{fig:u2} shows the optimal energy-error
tradeoff obtained with the algorithm of Prop. \ref{thm:algorithm}.

\subsection{Discrimination of passive devices with limited resources}
\label{sec:discr-pass-devic}

Here we consider the minimum energy discrimination that makes use of
coherent input states $\ket{\alpha_i}$ and homodyne detections
$X_{\varphi_i}$ to discriminate a single use of a $n$-modes passive
linear optical device randomly chosen in the set $\{I,U\}$ with equal
probabilities.
\begin{align}
  \begin{aligned}
    \Qcircuit @C=0.7em @R=1em { \prepareC{\alpha_1} & \qw & \multigate{2}{U_x}
      & \qw & \measureD{X_{\varphi_1}} \\ \prepareC{\alpha_2} & \qw &
      \ghost{U_x} & \qw & \measureD{X_{\varphi_2}} \\ \prepareC{\alpha_3} &
      \qw & \ghost{U_x} & \qw & \measureD{X_{\varphi_3}} }
  \end{aligned} \quad .
\end{align}
If we consider coherent input states $\ket{\alpha_i} $ on
mode $i$ the global input state is $\ket{\xi} =
\bigotimes_i\ket{\alpha_i}$ which corresponds to an energy value
\begin{align}
  \label{eq:cohenergy}
  E(\xi) := \bra{\xi}E\ket{\xi} = \frac12 + \bra{\xi}N\ket{\xi} = \frac12 +
  \sum_i|\alpha_i|^2
\end{align}
Since for any passive linear device $V$ we have that $V
\bigotimes_i\ket{\alpha_i} = \bigotimes_i\ket{\beta_i}$ with $\ket{\beta_i}$
are coherent state, we can assume $U$ to be diagonal, i.e.  $U = \sum_i
e^{i\delta_i a_i^\dagger a_i}$.  The evolution of $\ket{\xi}$ under the action of
$U$ is then given by
\begin{align}\label{eq:cohevolution}
  U \ket{\xi} = \bigotimes_i \ket{e^{i\delta_i}\alpha_i}.
\end{align}

A quantum homodyne detection $X_\varphi$ is described \cite{VW06,TS04,BW96} by
the POVM $ \{\ketbra{x,\varphi}{x'gamma} \}$ where $\ket{x,\varphi}$ are the
eigenstate of the quadrature $e^{i \varphi}a + e^{- i \varphi}a^\dagger$.  The
probability of outcome $x$ when the system is prepared in a coherent state
$\ket{\alpha}$, $\alpha = e^{i \phi_\alpha} |\alpha|$ is given by the Gaussian
\begin{align}
  \label{eq:gaussian}
 p_\varphi(x|\alpha) = |\braket{\alpha}{x,\varphi}|^2 = \frac{1}{\sqrt{2\pi}}
 e^{-\frac12(x-2|\alpha| \cos(\varphi + \phi_\alpha))^2}.
\end{align}
We notice that $p_\varphi(x|\alpha)$ depends on $\varphi$ only trough the sum
$\varphi + \phi_\alpha$. We can then fix $\varphi = 0$ and vary only the
$\alpha_i$. conditional probabilities of outcome ${\bf x} = (x_i)$ given $I$
or $U$ are $n$-dimensional Gaussians, namely
\begin{align}\label{eq:gaussians}
  & p({\bf x}|I) = (2\pi)^{-n/2} e^{|{\bf{x}} - {\bf{v_0}}|^2/2
  },\nonumber\\ & p({\bf x}|U) = (2\pi)^{-n/2} e^{|{\bf{x}} -
    {\bf{v_1}}|^2/2}, \nonumber\\ &{\bf{v_0}} = (2 \re \alpha_i), \qquad
  {\bf{v_1}} = (2 \re e^{i \delta_i}\alpha_i),
\end{align}

Any classical postprocessing of the outcome ${\bf x}$ can be described by a
function $q(X|{\bf x})$ that evaluates to $1$ if one guess the unitary $X$
from outcome ${\bf x}$, and to $0$ otherwise, with $X=I,U$.  The probability
of error is given by
\begin{align}\label{eq:coherror}
  P_e(\xi) = \frac12 \int \!\! d {\bf x} \, p({\bf x}|I)q(U|{\bf x}) 
  + p({\bf x}|U)q(I|{\bf x}),
\end{align}
and thus the optimal
postprocessing is
\begin{align}\label{eq:postproc}
  q(X|{\bf x}) = \left\{ \begin{array}{ll} 1 & \textrm{if } p({\bf x}|X) \ge
    p({\bf x}|Y)\\ 0 & \textrm{if } p({\bf x}|X) < p({\bf x}|X) \end{array}
\right. .
\end{align}
Inserting Eq. \eqref{eq:postproc} and Eq. \eqref{eq:gaussians}
into the expression \eqref{eq:coherror}, the probability of error
becomes
\begin{align}
  \label{eq:errorcoherent}
  P_e(\xi) = \frac12 + \frac{(2\pi)^{-n/2}}2 \int_{A} \!\! d {\bf x} \,
  e^{-\frac{|{\bf{x}} - {\bf{v_0}}|^2}2 } - e^{-\frac{|{\bf{x}} -
      {\bf{v_1}}|^2}2},
\end{align}
where we defined the set 
\begin{align}
  \label{eq:halfspace}
  A = \{ {\bf x} \mbox{ s.t. }  |{\bf{x}} - {\bf{v_0}}|^2 \geq |{\bf{x}} -
  {\bf{v_1}}|^2 \}.
\end{align}
Within this framework it is more convenient to fix the amount of energy, that
is the average number of photons $\eta$, and find the input state
$\ket{\xi^*}$ that minimizes the probability of error in the discrimination,
i.e.
\begin{align}
  \label{eq:2}
  \ket{\xi^*} = \arg \min_{\bra{\xi}N\ket{\xi}=\eta} P_e(\xi).
\end{align}

With a little machinery it is possible to prove that $P_e(\ket{\xi})$ is a
non-increasing function of $|{\bf{v_0}}-{\bf{v_1}}|^2$ and then the
minimization of $P_e(\ket{\xi})$ can be rephrased as a maximization of
$|{\bf{v_0}}-{\bf{v_1}}|^2$. We have then
\begin{align}
  |{\bf{v_0}}-{\bf{v_1}}|^2 &= 4\sum_i [\re(\alpha_i) - \re(e^{i
    \delta}\alpha_i)]^2 = \nonumber \\ &= 4\sum_i [(\cos(\phi_i) - \cos(\phi_i +
  \delta_i) )|\alpha_i|]^2 \leq\nonumber\\ &\leq 4\sum_i [2 \sin(\delta_i/2)
  |\alpha_i|]^2\le 16\sin^2(\delta/2) \eta \label{eq:coherrorbound}
\end{align}
where $\delta^* := \arg\max_{\delta_i} |\delta_i|$, and $i^*$ labels the
corresponding mode. The bounds in Eq. \eqref{eq:coherrorbound} are achieved
for
\begin{align}
  \ket{\xi^*} = \bigotimes_{i \neq i^*} \ket{0_i} \otimes \ket{\alpha_{i^*}^*},
\end{align}
where $\alpha^*_{i^*} = \sqrt{\eta}\exp(i\frac{\pi-\delta^*}{2})$. The
corresponding optimal discrimination strategy is
\begin{align}
  \begin{aligned}
    \Qcircuit @C=0.7em @R=1em {
      \prepareC{\alpha^*_{i^*}} & \qw &   \multigate{2}{U_x} & \qw & \measureD{X_0} \\
      \prepareC{0} & \qw & \ghost{U_x} & \qw &
      \measureD{I} \\
      \prepareC{0} & \qw & \ghost{U_x} & \qw &
      \measureD{I} }
  \end{aligned} \quad 
\end{align}
where $  \begin{aligned}
    \Qcircuit @C=0.7em @R=1em {& \measureD{I} \qw }
  \end{aligned} $
means that the corresponding mode is discarded.
With this choice of the input state the probability of error becomes
\begin{align}\label{eq:coherrtradeoff}
  P_e = \frac12\left(1 + \Phi\left(-2\sqrt{\eta}\sin\frac{|\delta^*|}2\right) -
  \Phi\left(2\sqrt{\eta}\sin\frac{|\delta^*|}2\right)\right),
\end{align}
where $\Phi(x) = \frac{1}{\sqrt{2\pi}}\int_{-\infty}^x dt \exp(-t^2/2)$ denotes
the normal cumulative distribution function.

\begin{figure}[htb]
  \includegraphics{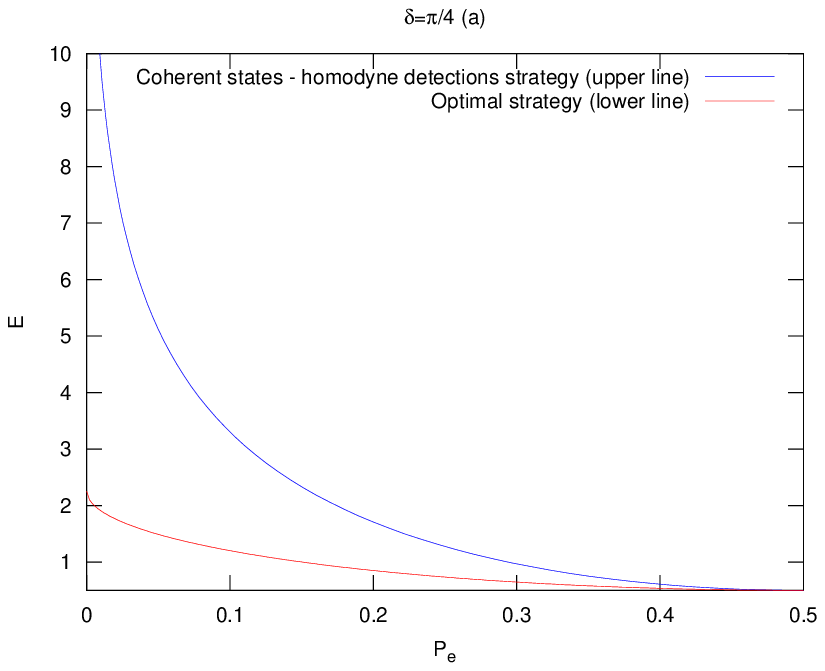}
  \includegraphics{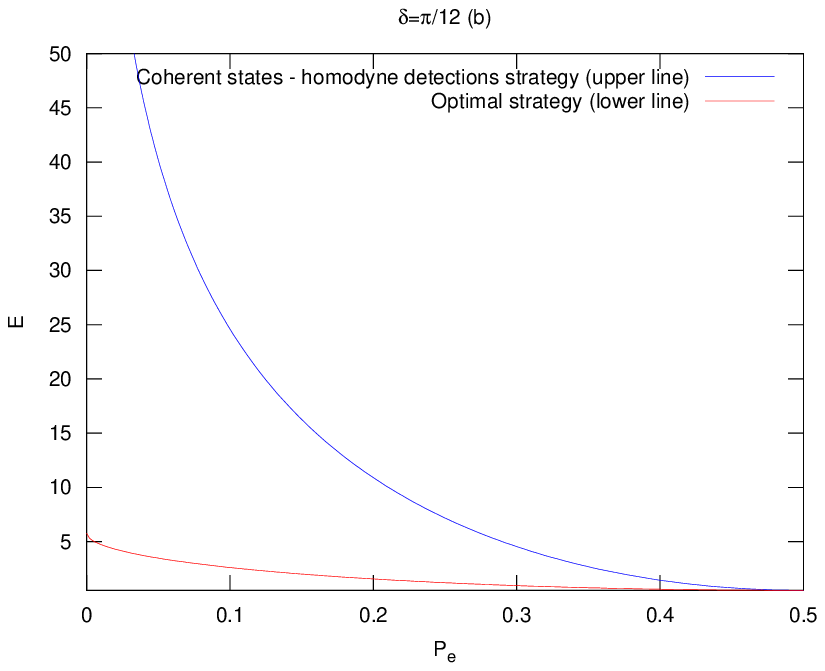}
  \caption{(Color online) Optimal tradeoff between the energy $E$ and the
    probability of error $P_e$ in the discrimination of $I$ and $U=\exp(i
    (\delta a_1^\dagger a_1 - \delta a_2^\dagger a_2))$ ($\delta=\pi/4$ in (a)
    and $\delta=\pi/12$ in (b)). The upper line represents the discrimination
    with coherent states and homodyne detections, while the lower line
    represents the optimal discrimination. Comparing (a) and (b), we notice
    that the improvement provided by the optimal strategy increases as
    $\delta$ decreases.}
  \label{fig:cstradeoff}
\end{figure}

From Eq. \eqref{eq:coherrtradeoff} one can obtain the tradeoff between the
energy and the probability of error, which is plotted in
Fig. \ref{fig:cstradeoff}, for some choices of $U_1$ and $U_2$.
If we consider the case in which we want to discriminate a $50/50$
beamsplitter from the identity, one can notice that, for $P_e=0.1$, the
coherent state - homodyne detection discrimination strategy requires a factor
of $\sim 4$ more photons that the optimal strategy. Moreover, this factor
increases as the two devices get closer, i. e. for small values of
$\delta$. For example, when $\delta=\pi/12$, the factor is $\sim 12$.
As expected, one notice that this factor increases when the probability of
error decreases.

\section{Conclusion}
\label{sec:conclusion}

In this paper we studied the energy-error tradeoff in the discrimination of
two linear passive optical devices. We derived the optimal strategy when no
restrictions in the input state and in the final measurement are assumed. It
is shown that the input state can be taken as pure and no ancillary modes are
needed. For the beamsplitter case the input state is proved to be a coherent
superposition of a NOON state $\ket{\phi_{n^*}}$ and the vacuum $\ket{00}$. It
is worth noting that the choice of $\ket{\phi_{n^*}}$ depends only upon the
reflectivity of the beamsplitter. We provided an iterative algorithm to solve
the problem in the general case. 

In Section \ref{sec:discr-pass-devic} we considered the practical scenario in
which one performs the discrimination with coherent input states and homodyne
detections. This strategy, when compared with the optimal one, turns out to be
largely suboptimal. For example, for a $50/50$ beamsplitter the energy
required for the discrimination when the probability of error is $0.1$ is $4$
times smaller with the optimal strategy than with the coherent-homodyne one.

\section*{Acknowledgments}

We thank Paolo Perinotti, Massimiliano F. Sacchi, and Michal Sedl\'{a}k for
useful suggestions and discussions. This work is supported by Italian Ministry
of Education through PRIN 2008 and the European Community through the COQUIT
and CORNER projects.

\end{document}

%% file: myqcircuit.tex
\usepackage[matrix,frame,arrow]{xy}
\usepackage{amsmath}
\newcommand{\bra}[1]{\left\langle{#1}\right\vert}
\newcommand{\ket}[1]{\left\vert{#1}\right\rangle}
\newcommand{\qw}[1][-1]{\ar @{-} [0,#1]}

\newcommand{\gate}[1]{*{\xy *+<.6em>{#1};p\save+LU;+RU **\dir{-}\restore\save+RU;+RD **\dir{-}\restore\save+RD;+LD **\dir{-}\restore\POS+LD;+LU **\dir{-}\endxy} \qw}

\newcommand{\measureD}[1]{*{\xy*+=+<.5em>{\vphantom{\rule{0em}{.1em}#1}}*\cir{r_l};p\save*!R{#1} \restore\save+UC;+UC-<.5em,0em>*!R{\hphantom{#1}}+L **\dir{-} \restore\save+DC;+DC-<.5em,0em>*!R{\hphantom{#1}}+L **\dir{-} \restore\POS+UC-<.5em,0em>*!R{\hphantom{#1}}+L;+DC-<.5em,0em>*!R{\hphantom{#1}}+L **\dir{-} \endxy} \qw}

\newcommand{\multimeasureD}[2]{*+<1em,.9em>{\hphantom{#2}}\save[0,0].[#1,0];p\save !C *{#2},p+LU+<0em,0em>;+RU+<-.8em,0em> **\dir{-}\restore\save +LD;+LU **\dir{-}\restore\save +LD;+RD-<.8em,0em> **\dir{-} \restore\save +RD+<0em,.8em>;+RU-<0em,.8em> **\dir{-} \restore \POS !UR*!UR{\cir<.9em>{r_d}};!DR*!DR{\cir<.9em>{d_l}}\restore \qw}

\newcommand{\multigate}[2]{*+<1em,.9em>{\hphantom{#2}} \qw \POS[0,0].[#1,0];p !C *{#2},p \save+LU;+RU **\dir{-}\restore\save+RU;+RD **\dir{-}\restore\save+RD;+LD **\dir{-}\restore\save+LD;+LU **\dir{-}\restore}
\newcommand{\ghost}[1]{*+<1em,.9em>{\hphantom{#1}} \qw}

\newcommand{\ustick}[1]{*!D!<0em,-.5em>=<0em>{#1}}

\newcommand{\Qcircuit}[1][0em]{\xymatrix @*=<#1>}

\newcommand{\pureghost}[1]{*+<1em,.9em>{\hphantom{#1}}}
\newcommand{\multiprepareC}[2]{*+<1em,.9em>{\hphantom{#2}}\save[0,0].[#1,0];p\save !C
  *{#2},p+RU+<0em,0em>;+LU+<+.8em,0em> **\dir{-}\restore\save +RD;+RU **\dir{-}\restore\save
  +RD;+LD+<.8em,0em> **\dir{-} \restore\save +LD+<0em,.8em>;+LU-<0em,.8em> **\dir{-} \restore \POS
  !UL*!UL{\cir<.9em>{u_r}};!DL*!DL{\cir<.9em>{l_u}}\restore}
\newcommand{\prepareC}[1]{*{\xy*+=+<.5em>{\vphantom{#1\rule{0em}{.1em}}}*\cir{l^r};p\save*!L{#1} \restore\save+UC;+UC+<.5em,0em>*!L{\hphantom{#1}}+R **\dir{-} \restore\save+DC;+DC+<.5em,0em>*!L{\hphantom{#1}}+R **\dir{-} \restore\POS+UC+<.5em,0em>*!L{\hphantom{#1}}+R;+DC+<.5em,0em>*!L{\hphantom{#1}}+R **\dir{-} \endxy}}

%% file: discoptdev.bbl
\begin{thebibliography}{}
\bibitem{Lloydscience} S. Lloyd, Science {\bf 321}, 1463 (2008) 
\bibitem{TEGGLMPS08} S. H. Tan, B. I. Erkmen, V. Giovannetti, S. Guha,
  S. Lloyd, L. Maccone, S. Pirandola, J. H. Shapiro, Phys. Rev. Lett. {\bf
    101}, 253601 (2008).
\bibitem{Lloydshapnewjphys} J. H. Shapiro and S. Lloyd, New J. Phys. {\bf 11}
  063045 (2009).
\bibitem{Aci01} A. Acin, Phys. Rev. Lett 87, 177901 (2001).
\bibitem{DLP02} G. M. D' Ariano, P. Lo Presti, and M. G. A. Paris, J. Opt. B
  {\bf 4}, 273 (2002).
\bibitem{DFY07} R. Duan, Y. Feng, M. Ying, Phys. Rew. Lett. {\bf 98}, 100503
  (2007).
\bibitem{sedlakdiscrmin} M. Ziman and M. Sedl\'{a}k, Journal of Modern Optics,
  {\bf 57}, 253 (2010).
\bibitem{collinsdiscrimin} D. Collins, Phys. Rev. A {\bf 81}, 052323 (2010).
\bibitem{note} The scenario in which multiple uses are considered involves the
  optimization over all the possible strategies (parallel, sequential,
  hybrid), and will be discussed in a future work.
\bibitem{Hel76} C. W. Helstrom, {\em Quantum Detection and Estimation Theory}
  (Academic Press, New York, 1976).
\bibitem{Pir11} S. Pirandola, Phys. Rev. Lett. {\bf 106}, 090504 (2011).
\bibitem{Nai11} R. Nair, arXiv:quant-ph/1105.4063v2 (2011).
\bibitem{IPP10} C. Invernizzi, M. G. A. Paris, and S. Pirandola,
  arXiv:quant-ph/1011.2785v4 (2010).
\bibitem{VW06} W. Vogel and D. G. Welsch, {\em Quantum Optics} (Wiley-VCH,
  Weinheim, 2006).
\bibitem{leonhardt03} U. Leonhardt, Rept. Prog. Phys. {\bf 66}, 1207 (2003).
\bibitem{NC00} I. L. Chuang, M. A. Nielsen, {\em Quantum Information and
  Communication} (Cambridge, Cambridge University Press, 2000).
\bibitem{San89} B. C. Sanders, Phys. Rev. A {\bf 40}, 2417 (1989).
\bibitem{BV04} S. Boyd and L. Vandenberghe, {\em Convex Optimization}
  (Cambridge, New York, 2004).
\bibitem{Arf85} G. Arfken, {\em Mathematical Methods for Physicists} (Academic
  Press, Orlando, 1985).
\bibitem{TS04} T. Tyc, B. C. Sanders, J. Phys. A {\bf 37}, 7341 (2004).
\bibitem{BW96} K. Banaszek, K. Wodkiewicz, Phys. Rev. A {\bf 55}, 3117 (1997).
\end{thebibliography}
